\documentclass[%
superscriptaddress,
showpacs,preprintnumbers,
amsmath,amssymb,
aps,
prl,
twocolumn,
notitlepage
]{revtex4-1}

\usepackage{microtype}
\usepackage{graphicx}   
\graphicspath{{fig/}}   
\usepackage{import}
\usepackage{dcolumn}
\usepackage{bm}
\usepackage{hyperref}

\usepackage{siunitx}
\usepackage{chemformula}

\newcommand*{\ii}{\mathrm{i}}
\newcommand*{\ee}{\mathrm{e}}
\newcommand*{\Alox}{\ch{AlO_x}}

\begin{document}


\title{Rabi oscillations in a superconducting nanowire circuit}


\author{Yannick Sch\"on}
\affiliation{Physikalisches Institut, Karlsruhe Institute of Technology, 76131 Karlsruhe, Germany}
\author{Jan Nicolas Voss}
\affiliation{Physikalisches Institut, Karlsruhe Institute of Technology, 76131 Karlsruhe, Germany}
\author{Micha Wildermuth}
\affiliation{Physikalisches Institut, Karlsruhe Institute of Technology, 76131 Karlsruhe, Germany}
\author{Andre Schneider}
\affiliation{Physikalisches Institut, Karlsruhe Institute of Technology, 76131 Karlsruhe, Germany}
\author{Sebastian T. Skacel}
\affiliation{Physikalisches Institut, Karlsruhe Institute of Technology, 76131 Karlsruhe, Germany}
\author{Martin P. Weides}
\affiliation{Physikalisches Institut, Karlsruhe Institute of Technology, 76131 Karlsruhe, Germany}
\affiliation{School of Engineering, University of Glasgow, Glasgow, G12 8QQ, United Kingdom}
\author{Jared H. Cole}
\affiliation{Chemical and Quantum Physics, School of Science, RMIT University, Melbourne, Victoria, 3001, Australia}
\author{Hannes Rotzinger}
\email{rotzinger@kit.edu}
\affiliation{Physikalisches Institut, Karlsruhe Institute of Technology, 76131 Karlsruhe, Germany}
\affiliation{Institute for Quantum Materials and Technologies, Karlsruhe Institute of Technology, 76131 Karlsruhe, Germany}
\author{Alexey V. Ustinov}
\affiliation{Physikalisches Institut, Karlsruhe Institute of Technology, 76131 Karlsruhe, Germany}
\affiliation{National University of Science and Technology MISIS, Moscow 119049, Russia}
\affiliation{Russian Quantum Center, Skolkovo, Moscow 143025, Russia}

\makeatletter
\def\Dated@name{Revised: }
\makeatother
\date{\today}


\begin{abstract}

We investigate the circuit quantum electrodynamics of anharmonic superconducting nanowire oscillators. The sample circuit consists of a capacitively shunted nanowire with a width of about 20\,nm and a varying length up to 350\,nm, capacitively coupled to an on-chip resonator. By applying microwave pulses we observe Rabi oscillations, measure coherence times and the anharmonicity of the circuit. Despite the very compact design, simple top-down fabrication and high degree of disorder in the oxidized (granular) aluminum material used, we observe lifetimes in the microsecond range.

\end{abstract}



\maketitle



\section{Introduction}

Quantum electrodynamics experiments with superconducting circuits (cQED) usually feature one or more Josephson tunnel junctions embedded in a circuit. Such circuits often feature a non-linear inductive response of the Josephson junction, leading to discrete, non-equidistant, energy levels \cite{Makhlin2001, Devoret2004}. Alternative approaches to the realization of non-linear elements are, for example, explored using hybrid quantum systems \cite{Larsen2015, Wang2019}.

In this paper, we report on a quantum circuit which employs a superconducting nanowire as a non-linear element. Using the powerful cQED approach, material properties arising at the nanometer scale are studied by directly observing measures like inductance, non-linearity, or coherence. We demonstrate that such a simple circuit has a rather long ($\sim\si{\micro\second}$) excited state life time.

As a material we use oxidized (granular) aluminum (\Alox) \cite{Rotzinger2016} which has recently been introduced into large impedance quantum circuit applications \cite{Grunhaupt2019}, due to its low loss properties at microwave frequencies, also in the single photon regime \cite{Grunhaupt2018}. \Alox\ films consist of nanometer sized aluminum grains (on average of about 4\,nm in diameter) embedded in an insulating aluminum-oxide matrix (see e.g. supplementary material in \cite{Rotzinger2016}). The inter-grain tunnel barriers lead to a sheet resistance of up to few k$\Omega$ which can be controlled during the film growth by adjusting the oxygen partial pressure.

In general, a wire made from \Alox\ can be seen as a series of conducting grains separated by insulating barriers. If the wire width is comparable with the aluminum grain size, its superconducting properties resemble the behavior of a disordered chain of Josephson weak links \cite{Maleeva2018}. Throughout this paper, the studied wires have a width $w$ of about 20\,nm.

The nanoscale structure gives rise to the main difference between the nanowire and the Josephson junction. While the lumped Josephson tunnel barrier of a traditional quantum circuit has a sinusoidal current-phase relation, the phase drop along an \Alox\ nanowire is distributed over many nanometers and has a more linear current-phase relation involving hundreds of microscopic Josephson weak links. Making use of this non-linearity for a new type of quantum device poses an intriguing challenge.

In terms of electric loss, nanowires can be advantageous since an applied voltage drops over many junctions and thus the local electric fields are substantially reduced compared to those in a single junction. Therefore, two-level defects present in the vicinity of the local barriers should couple only weakly to the electrical field of such a circuit \cite{Grabovskij2012, Bilmes2017}. However, the influence of unpaired spins in the nanoscale granular material \cite{Bachar2013, Barone2018}, e.g. due to the parity effect in the grains, is still to be understood. 

From the perspective of a nanowire, two regimes have to be considered. A long and very thin wire with a high normal state resistance ($\gg\SI{50}{\kilo\ohm}$) undergoes a transition to an insulating state at low temperatures ($T<1$\,K). Here, due to quantum fluctuations, the superconducting phase is not well defined and, consequently, the electric transport for excitations below the superconducting gap is suppressed. This corresponds to the quantum phase slip regime which has been investigated in the context of homogeneous wires \cite{Lau2001, Mooij2005, Mooij2006, Astafiev2012}. The nanowires considered in this paper instead have a lower normal state resistance and, due to the short coherence length $(\xi = (8\pm0.4)\,\si{\nm} \simeq w$ \cite{Voss2019}), the superconducting phase is well defined in the wire. This means that Cooper pairs can tunnel coherently along the wire up to its critical current $I_c$. Ignoring the local structure of the wire, this behavior is rather well described by a mean field approach of the Kulik-Omelyanchuk (KO) model \cite{Kulik1975, Voss2019}, which relates $I_c$ to the wire's superconducting gap and normal state resistance.

The measurement of quantum coherence in such one dimensional systems constitutes an attractive goal. Transferring this to the high resistance regime could allow for a distinction of dissipative and dissipationless phase slips. Additionally this would give access to intrinsic dynamics and their contributions to the anharmonicity in disordered one dimensional systems.

\section{Results and Discussion}

\subsection{Theory and Design}

An excitation of the capacitively shunted nanowire having the energy $E=\hbar \omega_{01}$ leads to a current $I \simeq \sqrt{2E/L}$, where $L$ is the total inductance of the circuit. The non-linearity of the nanowire depends on current $I$ to the lowest order as $(I/I_c)^2$ \cite{Anlage1989}. Therefore, in order to operate the nanowire in a sufficiently anharmonic qubit-like regime, it is useful to keep $I$ as close to  $I_c$ as possible. This can be achieved by two basic approaches: Either designing the circuit for a maximal $I$ (low total $L$) or operating it with a comparably large inductance $L$ at a much smaller $I_c$ (or a combination of both). The first option aims at minimizing the amount of high kinetic inductance material, for instance, by using a conventional low kinetic inductance superconductor like aluminum for the capacitive parts. This approach has the advantage that the geometric requirements for the nanowire are less stringent \cite{Vijay2009, Levenson-Falk2011}. This, however,  leads to a relatively large electric field across the nanowire and thus to potentially increased dielectric losses. We have chosen the second option and fabricated the whole circuit containing a low $I_c$ nanowire from a material with a rather large kinetic inductance, thus sacrificing the strength of non-linearity. The circuit design approach is based on the well known capacitively shunted Josephson junction (transmon-like) qubit \cite{Koch2007}, here with substantial differences due to the added large additional kinetic inductance (in the order of a few nH). The large impedance $Z = \sqrt{L/C} \simeq 1\,\mathrm{k}\Omega$ of the circuit and the vacuum impedance are thus mismatched to reduce the effects of environmental noise.

The circuit is described by a Hamiltonian similar to the capacitively shunted Josephson junction qubit \cite{Koch2007, Schreier2008}, namely  $\mathcal{H}_\mathrm{T} = 4E_\mathrm{C}(\hat{n}-n_\mathrm{g})^2-E_\mathrm{J}\cos\hat{\varphi}$ with charging energy $E_\mathrm{C}=e^2 / 2C$ and Josephson coupling energy $E_J=\Phi_0 I_\mathrm{c} / 2\pi$. In a nanowire, the current-phase relation is non-sinusoidal and has a form, which we denote as $I\sim{f'}(\varphi)$. While its exact form is unknown, from the studies of superconducting weak links we anticipate it to be a function shaped between sine and sawtooth form (see e.g. \cite{Christiansen1971, Likharev1979, Wilkinson2018}). We assume orders of magnitude larger critical currents in the capacitive part of the circuit and thus neglect the non-linearity of the kinetic inductance contribution there.  We split the Josephson coupling term in the Hamiltonian into a linear part and the contribution of an effective junction (circuit design and diagram in Supplementary Note and Figure 1). Following the circuit quantization \cite{Devoret1997} approach, a Hamiltonian can be written as $ \mathcal{H} = 4E_\mathrm{C}(\hat{n}-n_\mathrm{g})^2-\tilde{E}_\mathrm{J}f(\hat{\varphi})+\tilde{E}_\mathrm{L}\hat{\varphi}^2$. Here $\tilde{E}_\mathrm{J} = 6 E_\mathrm{J} E_\mathrm{L}^2 / (6E_\mathrm{J}+2E_\mathrm{L})^2$ and $\tilde{E}_\mathrm{L} = 9 E_\mathrm{J}^2 E_\mathrm{L} / (6E_\mathrm{J}+2E_\mathrm{L})^2$. $E_\mathrm{L}$ is the inductive energy $(\Phi_0/2\pi)^2/2L$. This is valid in the limit of $E_\mathrm{L}$ being small compared to all other energy terms \cite{Braumuller2016}. To achieve a sufficient non-linearity of the circuit, the wire has to be long enough as to ensure it dominates the inductance of the system. It remains, however, to be studied how the wire length influences the non-linearity of the current-phase relation.

\begin{figure}
	\centering
	\includegraphics[width=\columnwidth]{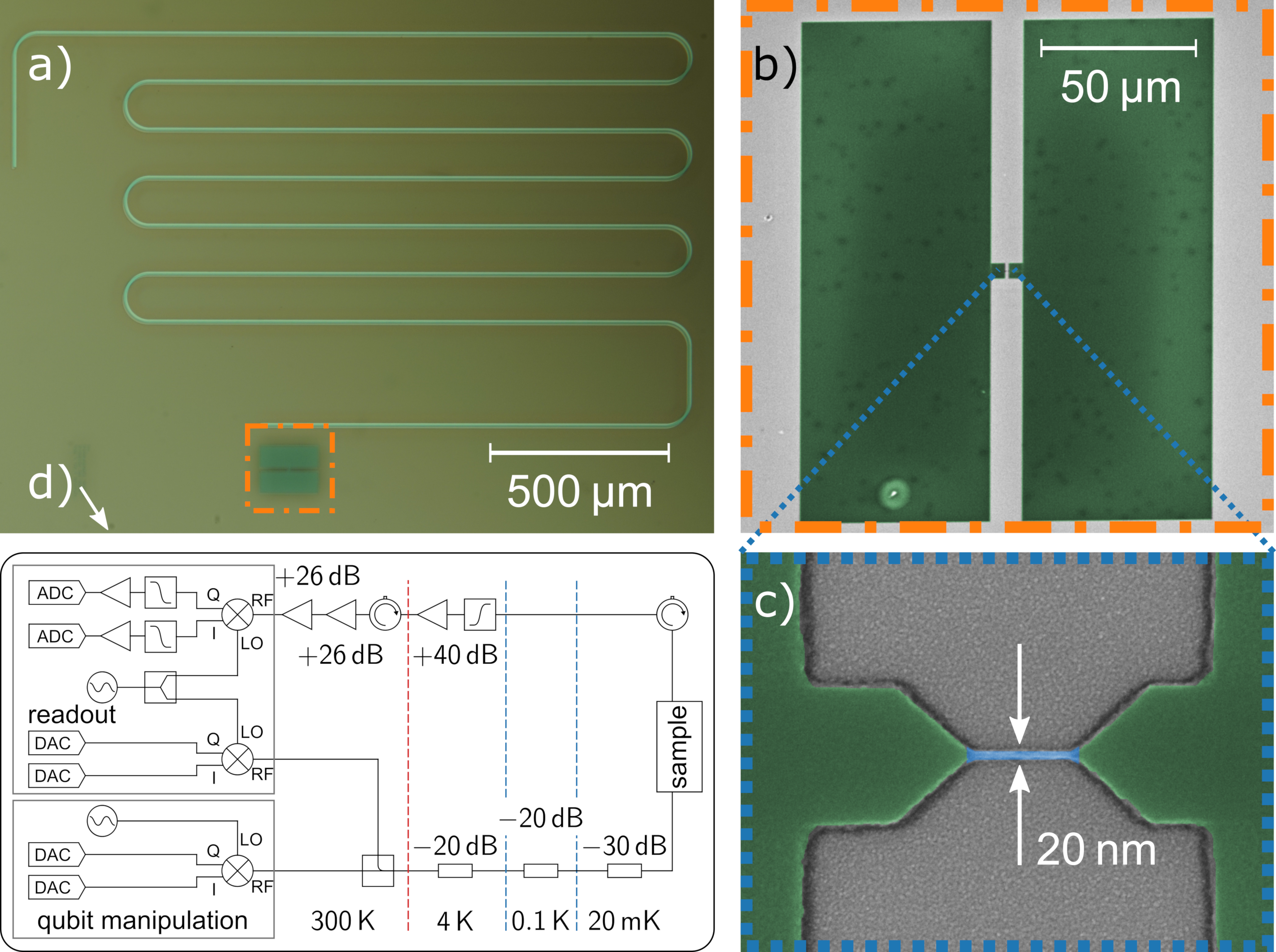}
	\caption{\textbf{Sample and setup. }\textbf{(a)} Optical microscopy photograph of a granular aluminum oxide nanowire oscillator (orange border) capacitively coupled to an aluminum resonator (circuit diagram in Supplementary Figure 1). \textbf{(b)} Scanning electron microscope (SEM) picture of one nanowire oscillator (colored in green). The large $60 \times \SI{160}{\um\squared}$ pads constitute the capacitance. Incremental constrictions lead to the nanowire. \textbf{(c)} Colored SEM closeup of the nanowire (blue) shunting the two capacitive pads. \textbf{(d)} Cryogenic microwave measurement setup.}
	\label{fig:sample}
\end{figure}

The nanowire is capacitively shunted by two rectangular $60 \times \SI{160}{\micro\m}$ capacitor pads, with a distance of \SI{10}{\um} (see Fig. \ref{fig:sample}). As obtained by analytical calculation \cite{Gevorgian2001} and simulation (ANSYS Maxwell), the pads constitute a $(22\pm2)\,\si{\femto\farad}$ capacitance. The studied three nanowire circuits have varied wire lengths of $50$, $100$ and $\SI{350}{\nm}$ and are coupled to individual readout resonators at 6.85, 6.10 and \SI{4.99}{\GHz}. For a summary of the sample parameters see Table \ref{tab:overview}.

\begin{table*}
	\caption{\textbf{Properties of the sample circuits.} The wire length is denoted by $l_\mathrm{W}$. $R_\mathrm{W}$ is the room temperature resistance of the wire estimated from the total circuit resistance. The readout resonator frequencies $f_\mathrm{r}$ and transition frequencies $f_{01}$ are obtained from spectroscopy. $f_\mathrm{calc.}$ corresponds to the harmonic estimate of the circuits resonance. The lifetimes are obtained by time-domain measurements.\label{tab:overview}}
	\begin{ruledtabular}
		\begin{tabular}{l l l l l l l l}
			Sample & $l_\mathrm{W}$ (\si{\nm}) & $R_\mathrm{W}$ (\si{\kilo\ohm}) & $f_\mathrm{r}$ (\si{\GHz}) & $f_{01}$ (\si{\GHz}) & $f_\mathrm{calc.}$ (\si{\GHz}) & Lifetime (\si{\micro\s}) \\ \hline \noalign{\smallskip}
			S1 & \tablenum{350} & $<7$ & 4.99 & 5.47 & $5.2\pm0.3$ & $3.4\pm0.1$ ($0.8\pm0.2$ $T_2$) \\
			S2 & \tablenum{50} & $7\pm1$ & 6.85 & 8.50 & $9.2\pm0.4$ & $4.0\pm0.1$ \\
			S3 & \tablenum{100} & $10\pm1$ & 6.10 & 7.93 & $8.3\pm0.4$ & $3.4\pm0.1$  \\
		\end{tabular}
	\end{ruledtabular}
\end{table*}

\subsection{Measurements}

\begin{figure}
	\includegraphics[width=0.9\columnwidth]{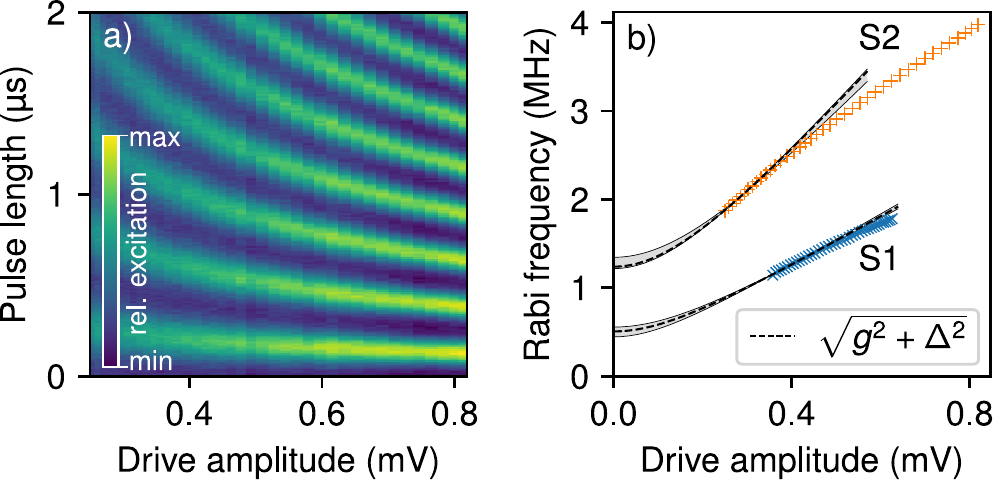}
	\caption{\textbf{Rabi oscillation power dependence. }\textbf{(a)} Rabi oscillations measured in sample S2 with varied drive amplitude. \textbf{(b)} The fitted Rabi frequencies measured in S1 as well as S2 fit to the two-level Rabi behavior until they deviate at frequencies exceeding the anharmonicity. The values for the drive detuning $\Delta$ obtained from the fit are $\Delta^\mathrm{fit}_\mathrm{S1}=(0.5\pm0.1)\,\si{\MHz}$ and $\Delta^\mathrm{fit}_\mathrm{S2}=(1.3\pm0.1)\,\si{\MHz}$. Errorbands include a variation of the number of points fitted between 6 and 16. Compared with spectroscopy the Rabis were driven with $\Delta_\mathrm{S1}=(0.8\pm0.5)\,\si{\MHz}$ and $\Delta_\mathrm{S2}=(1.8\pm1)\,\si{\MHz}$ respectively. In sample S1, a smaller anharmonicity agrees with a lower normal state wire resistance (Table \ref{tab:overview}). 
		\label{fig:fofA}}
\end{figure}

\begin{figure}
	\centering
	\includegraphics[width=0.8\columnwidth]{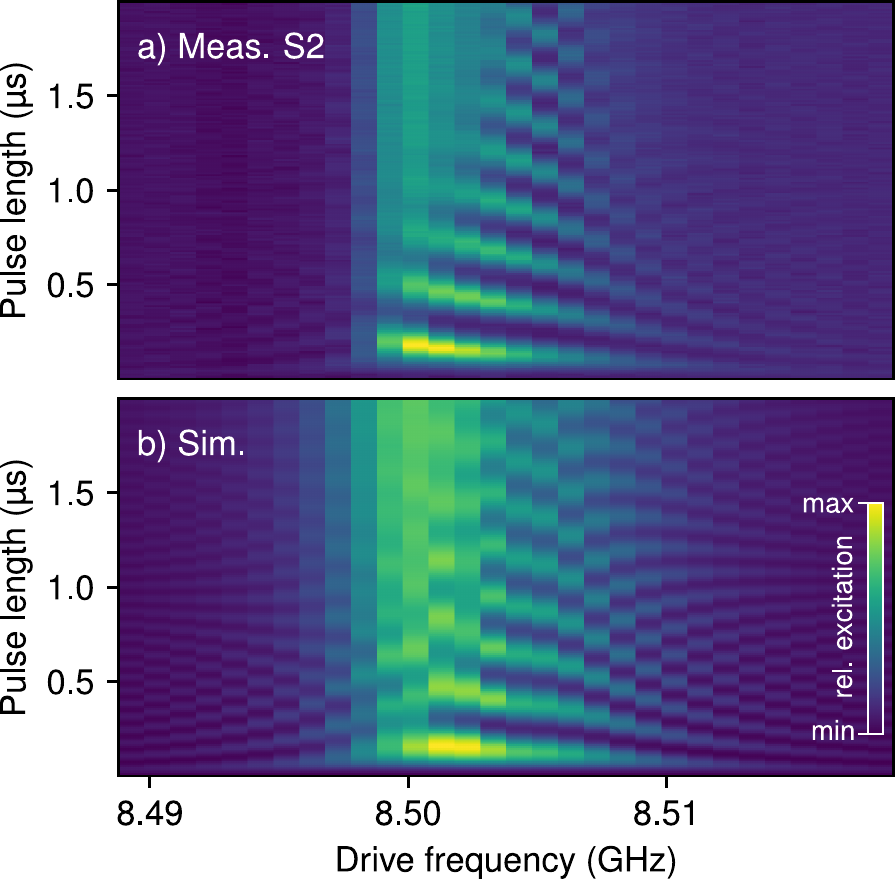}
	\caption{\textbf{Rabi oscillation frequency dependence. }\textbf{(a)} Rabi oscillations measured in S2 (\SI{50}{\nm} long wire) in a span of \SI{30}{\MHz} around the low-power transition frequency of \SI{8.504}{\GHz}. The distinct asymmetry can be attributed to higher level excitations and a relative anharmonicity between 1 to \SI{2}{\MHz}. \textbf{(b)} Result of the numerical Floquet matrix calculation of the density matrix master equation for an anharmonic oscillator with \SI{1.3}{\MHz} anharmonicity. This agrees with an estimation from Fig. \ref{fig:fofA}.}
	\label{fig:roff50}
\end{figure}

Three anharmonic oscillators with varying wire length were measured using a dispersive readout scheme \cite{Blais2004}. Transition frequencies $f_{01}$ were determined spectroscopically by observing the shift of the readout resonator's frequency induced by the second drive tone. The circuit's inductance can be derived from its normal conductive resistance as $ L = 0.18 \hbar R / (k_\mathrm{B}T_\mathrm{c}) $, where $T_\mathrm{c}$ is about \SI{1.8}{\kelvin} \cite{Rotzinger2016}. With the capacitance $C$, a harmonic approximation for the circuit's frequency is $f_\mathrm{calc.}=1/(2\pi\sqrt{LC})$. While this estimate agrees with the measured transition frequency of sample S1 (\SI{350}{\nm} long wire), for samples S2 (\SI{50}{\nm}) and S3 (\SI{100}{\nm}) the measured transition frequency is lower than the estimate. The later two samples also exhibit higher room temperature wire resistances and a higher wire to pad resistance ratio (Table \ref{tab:overview}).

A continuous microwave tone on resonance with a system's transition frequency results in Rabi oscillations between its ground and excited states. These oscillations are recorded using pulsed time-domain measurements. Figure \ref{fig:fofA}a depicts the time evolution of the circuit excitation in dependence of the drive amplitude, measured in sample S2. Brighter colors correspond to the system being excited. The microwave drive amplitude has been calibrated after the room temperature part of the microwave setup (Fig. \ref{fig:sample}) with a spectrum analyzer. This compensates for the nonlinearities of the IQ mixers. In case of a single two-level transition coupled to a radiative field, the oscillation frequency $\Omega$ is expected to depend linearly on the coupling $g$, which is proportional to the driving field amplitude. An additional small detuning $\Delta$ between the drive and transition frequencies results in the form $\Omega=\sqrt{g^2+\Delta^2}$ \cite{Nielsen}. The Rabi frequency $\Omega$ is extracted from damped sine fits and displayed in Fig. \ref{fig:fofA}b. In both depicted oscillators S1 (\SI{350}{\nm} long wire) and S2 (\SI{50}{\nm} long wire), $\Omega$ fits to the two-level Rabi behavior until it starts to deviate toward higher frequencies. This deviation is expected in the presence of higher levels when $\Omega$ exceeds the circuit's anharmonicity, given by the difference between the lowest energy level separations \cite{Claudon2004, Dutta2008}. In this regime, multi-photon transitions to higher levels occur with a higher rate, thus reducing the power proportion driving the main transition.

Direct spectroscopy of multi-photon transitions to higher levels requires increasing drive powers \cite{Braumueller2015}. In the case of a small anharmonicity, the resulting broadening of the fundamental resonance line makes multi-photon resonances difficult to resolve (Supplementary Note 2). However, by observing Rabi oscillations with varied drive frequency, effects of higher transitions manifest themselves in a distinct asymmetry (Fig. \ref{fig:roff50}a). At frequencies above the main transition, the Rabi oscillation frequency $\Omega$ increases as its amplitude decreases as expected from off-resonant driving. Toward lower frequencies however, $\Omega$ continues to decrease until the oscillation breaks down. Also the oscillation amplitude increases and the equilibrium excitation for long driving rises. These effects agree with a system exhibiting a small but non-vanishing, negative anharmonicity between 1 to \SI{2}{\MHz}, allowing the drive to excite the fundamental transition and transitions to higher levels.

The presented interpretation of the above described effects is supported by numerical simulation of the Lindblad-GKS master equation \cite{Lindblad1976, Gorini1976}, see Supplementary Note 3. The system $\mathcal{H}_0=h(f_{01}+f_\mathrm{s}\sqrt{\epsilon^2/(\epsilon^2+\Delta^2)})\hat{n}-h  f_\mathrm{an}(\hat{n}^2-\hat{n})$ with the bosonic number operator $\hat{n}$  describes an oscillator with frequency $f_{01}$ and anharmonicity $f_\mathrm{an}$. An additional asymmetry extending toward the off-resonant regions in the measurement is accounted for by a slight shift $f_\mathrm{s}$ of the main transition in a region defined by a parameter for the width $\epsilon$ and reduced with the detuning $\Delta$ of the drive. This shift is expected due to an AC Stark effect \cite{Schneider2018}. In this Lindbladian \cite{Lindblad1976, Gardiner2004} approach, two decay channels with corresponding rates are assumed to account for energy dissipation and dephasing. Here we are specifically interested in the interplay between finite anharmonicity and multi-photon (strong driving) effects. Therefore a Floquet expansion \cite{Shirley1965, Bain2001} is used to solve for the time evolution in the strongly periodically driven system (see Supplementary Note 3).

To reproduce the measurement performed on sample S2 by the simulation shown in Fig. \ref{fig:roff50}b, the main transition frequency and excitation lifetime were taken from separate measurements (Supplementary Notes 2 and 3). The region around the main transition is well reproduced assuming the anharmonicity of $(1.5\pm0.3)\,\si{\MHz}$.

\begin{figure}
	\centering
	\includegraphics[width=0.9\columnwidth]{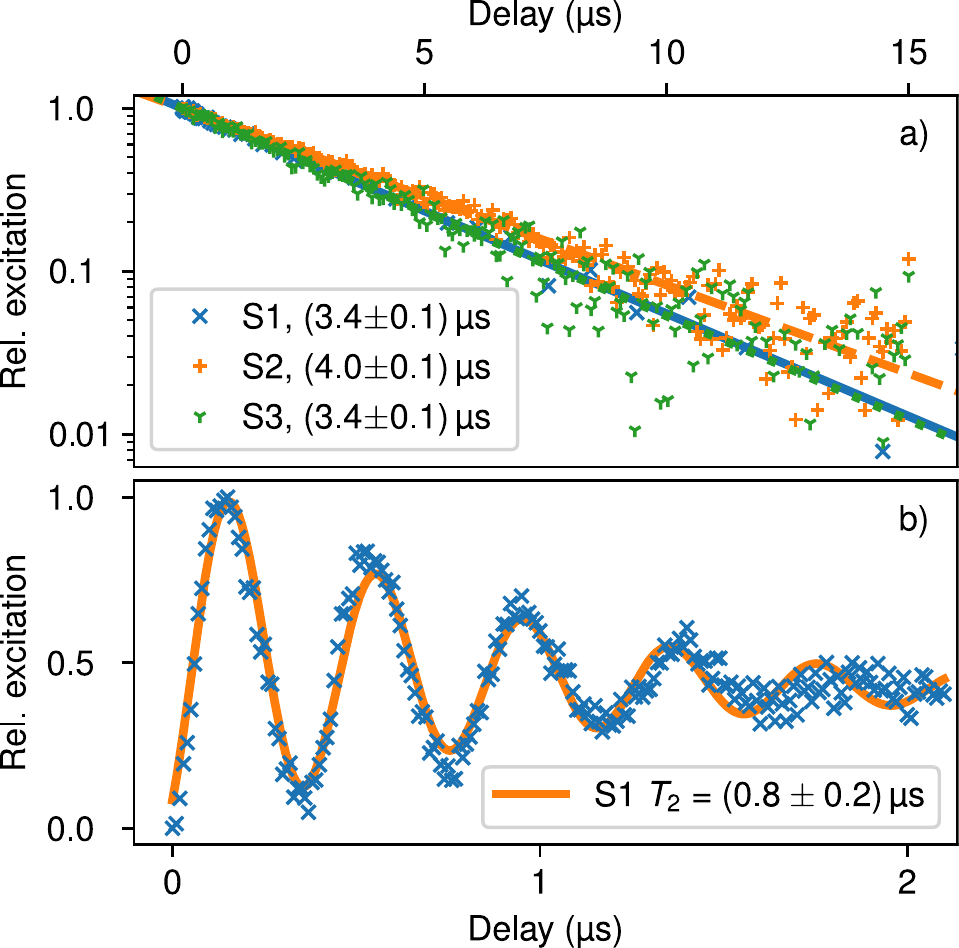}
	\caption{\textbf{Lifetime measurements. }\textbf{(a)} Excitation lifetimes $T_1$ measured in samples S1, S2 and S3 range between 3.4 and \SI{4}{\us}. \textbf{(b)} Ramsey fringes measured in S1 yield a dephasing time $T_2$ of ($0.8\pm0.2$)\,\si{\us}.}
	\label{fig:t1}
\end{figure}

Excitation decay times $T_1$ were measured by applying a microwave pulse $t_ {\pi}$ of half a Rabi period. Measurement of the excitation after a varied delay yields an exponential decay $\propto \ee^{-t/T_1}$. The $T_1$ times, measured in all samples, range between 3.4 and \SI{4}{\us} (Fig. \ref{fig:t1}a).

Excitation to a state on the equator of the Bloch sphere is possible by applying a $t_ {\pi/2}$-pulse. Off-resonant driving and varying the delay before applying the second $t_ {\pi/2}$-pulse results in Ramsey fringes (Fig. \ref{fig:t1}b). The decay of the fringes corresponds to the dephasing time $T_2$. Our measurement performed in sample S1 yielded a $T_2$ of \SI{0.82}{\us}.

We estimated single-mode Purcell loss $\gamma_\mathrm{P}=(g_\mathrm{r}/\Delta_\mathrm{r})^2\kappa$ \cite{Purcell1946, Houck2008} given by the coupling $g_\mathrm{r}=(17\pm2)\,\si{\MHz}$ to the readout resonator, its linewidth $\kappa=(1.5\pm0.2)\,\si{\MHz}$ and the frequency detuning $\Delta_\mathrm{r}=(487\pm1)\si{\MHz}$ for sample S1. The obtained Purcell lifetime limitation is $(85\pm14)\,\si{\micro\second} \gg T_1$. The large circuit impedance favors suppressing radiative loss. No direct correlation between transition frequency and lifetime has been observed. The contribution of dielectric loss due to randomly distributed two-level systems, which are always present in quantum circuits, can be more easily resolved with higher anharmonicity, as well as temperature or strain dependent studies \cite{Houck2008, Goetz2017, Muller2019}.

We expect that transitions to higher levels, which occur at frequencies close to the fundamental transition, contribute additional loss channels. This situation is facilitated by the finite bandwidth of the short control pulses. A rectangular \SI{100}{\ns} long $\pi$-pulse has a linewidth that can be estimated as $1/(2\pi\cdot\SI{100}{\ns})=\SI{1.6}{\MHz}$, thus being of the order of the observed anharmonicity. Optimized pulse shapes can help decrease the rate of unintended excitations \cite{Motzoi2009}. Excitations of higher transitions lead to additional frequency components modulating Ramsey fringes and thus make a precise measurement of the pure dephasing challenging. Also, charge noise due to a large charging energy $e^2/2hC$ of $(880\pm80)\,\si{\MHz}$ adds to the dephasing in the circuits. This contribution can be reduced together with a decrease of the total inductance by concentrating the circuit's inductance in the wire. A thus increased anharmonicity would allow for a distinction of the higher level contribution to the observed lifetimes and thus help anticipate the feasible performance of optimized \Alox\ nanowire qubits.

\section{Conclusion}

In this work, we demonstrated multi-level quantum dynamics in single-layer superconducting circuits. The anharmonic nature of these nanowire oscillators originates from nonlinear properties of the material they are made of, oxidized (granular) aluminum. These nanowire quantum circuits feature a simple, scalable fabrication process and are compact in design. The fabricated samples showed energy relaxation times $T_1$ of up to \SI{4}{\us} and a relatively small negative anharmonicity on the order of 1--\SI{2}{\MHz}. The measured characteristic Rabi patterns were replicated in numerical simulations. We thus demonstrated that nonlinearity and coherence times can be useful measures of the properties of granular materials structured into nanometer size circuits.

In the future, larger anharmonicities seem feasible through featuring nanowires with controllable smaller critical currents and reduced spread. To reduce the contribution of the capacitive paddles to the linear fraction of kinetic inductance, one can use additional shunts made of pure aluminum. These improvements would additionally enhance the circuit's usability as qubit. Studying of the system's Hamiltonian in dependence of the nanowire's geometry and resistance would be facilitated by direct spectroscopy of the level structure. This would shed further light on the physics of the non-linearity arising from the granular material in such systems.

\begin{small}

\section{Methods}

The disordered oxidized aluminum thin films are grown on a c-plane sapphire substrate by sputter deposition. Details of the process can be found in Ref. \cite{Rotzinger2016}. In addition, an in-situ resistance monitoring was used to allow for a precise control over the final sheet resistance of the film (k$\Omega$ range, \SI{20}{\nm} film thickness) \cite{Wildermuth2019}.

The nanowire circuits are defined in a single electron beam lithography step using a bilayer resist stack of hydrogen silsesquioxane (HSQ) and polymethylmethacrylat (PMMA) on top of the \Alox\ thin film. This approach has the advantage that the high resolution HSQ resist can be lifted off after the pattern transfer with an organic solvent. The pattern transfer into the \Alox\ thin film is carried out using an anisotropic oxygen and argon/chlorine reactive ion plasma. The aluminum resonators, feedline and backside metallization are deposited using optical lithography and the lift-off technique.

Due to the galvanic isolation of the nanowire circuits, we characterized the DC resistance at room temperature using needle probes. This method poses severe limitations on the DC measurement results since it is sensitive to the contact resistance and position of the needles on the thin film sample. Special care has to be taken to avoid scratches. Additionally, the wire resistance can be altered by too high probe currents during the resistance measurement \cite{Voss2019}.

Microwave spectroscopy of the circuits as well as pulsed time domain measurements are performed in a dilution cryostat at \SI{\sim20}{\milli\kelvin} (Fig. \ref{fig:sample}d).

\section{Data Availability}

The data that support the findings of this study are available from the corresponding author upon reasonable request.

\section{Code Availability}

Measurements and data analysis in the context of this paper were performed with the open source software toolkit Qkit (\href{https://git.io/qkit}{https://git.io/qkit}). The code for the numerical modeling is available from the corresponding author upon reasonable request.


\def\bibfont{\footnotesize}

\section{\label{sec:acknowledgment}{Acknowledgments}}
The authors want to thank J. Lisenfeld and I. Pop for fruitful discussions as well as L. Radtke and S. Diewald for assistance in sample fabrication and R. Gebauer for contributions to the measurement setup. This work was supported by the Initiative and Networking Fund of the Helmholtz Association, as well as by the Helmholtz International Research School for Teratronics (YS, JNV), the Landesgraduiertenf\"orderung of the federal state Baden-W\"urttemberg (MW) and the Carl Zeiss Foundation (AS). This work was partially supported by the Ministry of Education and Science of the Russian Federation in the framework of the Program to Increase Competitiveness of the NUST MISIS (contract No. K2-2017-081). JHC acknowledges support of the Australian Research Council Centre of Excellence funding scheme (CE170100039) and the NCI National Facility through the National Computational Merit Allocation Scheme. MPW acknowledges support of the European Research Council (ERC) under the Grant Agreement 648011. We acknowledge support by the KIT-Publication Fund of the Karlsruhe Institute of Technology.

\section{Author Contributions}

All authors contributed to discussions and interpretations of the results. The experiment was conceived by YS and HR. Measurements were performed by YS with support by AS and MPW. Fabrication was done and supported by YS, JNV, MW, STS, and HR. JHC additionally advised the numerical modeling. The project was supervised by HR and AVU.

\section{Competing Interests}

The Authors declare no competing interests.

\section{Additional Information}

This is a post-peer-review version of an article published in npj Quantum Materials. The final authenticated version is available online at: \href{https://doi.org/10.1038/s41535-020-0220-x}{doi.org/10.1038/s41535-020-0220-x}.

\end{small}


\clearpage
\renewcommand{\figurename}{Supplementary Figure}
\renewcommand\thefigure{\arabic{figure}}
\setcounter{figure}{0}
\renewcommand\thesection{\Alph{section}}
\renewcommand{\theequation}{Sup\arabic{equation}}


\section{Supplementary Material}

\subsection{Supplementary Note 1: Circuit Design}

The granular aluminum oxide nanowire oscillator can be described by an equivalent circuit diagram (Supplementary Figure 1) in terms of a series connection of capacitance, linear inductance and non-linear inductance. On chip, two pads ($60 \times \SI{160}{\um\squared}$) constitute the capacitance. These are shunted by the nanowire contributing the non-linear inductance. Since the whole oscillator is made from a high kinetic inductance material, all circuit elements contribute additional inductance that acts linearly due to their higher critical currents.

For readout and manipulation, the non-linear oscillator is capacitively coupled to a classical harmonic resonator. This is realized in a meandered aluminum $\lambda/2$ geometry. At low temperatures ($T\ll\SI{1}{\kelvin}$) the excitation of the nanowire oscillator influences the harmonic resonator's frequency. This is used in a dispersive manipulation and readout scheme \cite{Blais2004}.

The circuit is connected to the measurement setup by two port transmission through a microwave feedline.

\subsection{\label{SN2}{Supplementary Note 2: Spectroscopy}}

To detect transitions in the nanowire circuits, a dispersive two-tone spectroscopy scheme is used. At one fixed frequency, the resonance dip of the readout resonator is monitored while a second drive tone is swept over a given frequency range. The coupling between the readout resonator and the sample circuit results in a shift of the readout resonance frequency when a transition is excited \cite{Blais2004}. This shift leads to a change in the measured amplitude and phase. Supplementary Figure 2 depicts data of three sample circuits. The increased drive power required for direct spectroscopy of multi-photon transitions \cite{Braumueller2015} to higher levels result in a broadening of the fundamental transition. Thus, individual lines of higher levels are not resolved.

\begin{figure}
	\centering
	\includegraphics[width=0.6\columnwidth]{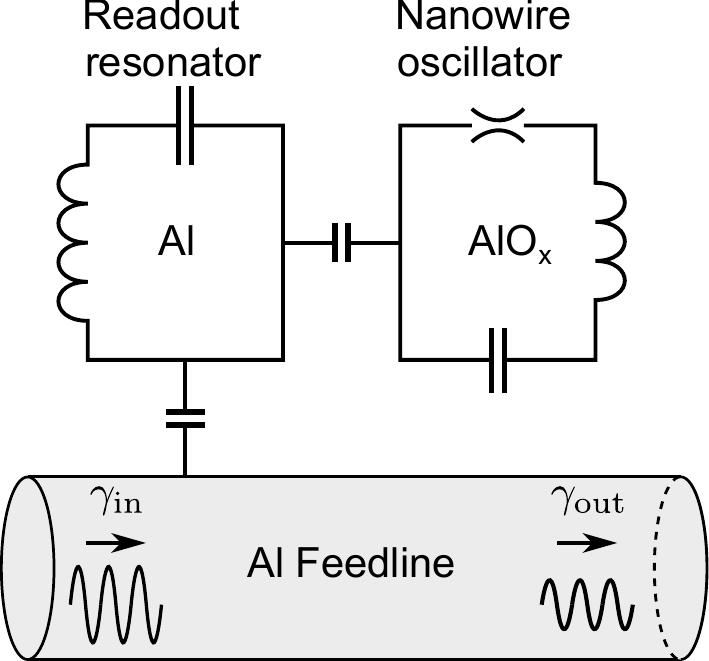}
	\caption{Schematic circuit diagram of the experiment. The granular aluminum oxide nanowire oscillator can be described as nonlinear LC-oscillator with its inductance split between the nonlinear contribution of the nanowire and the linear part in the capacitive paddles. It is coupled to a harmonic aluminum readout resonator with a resonance frequency depending on the circuit's state \cite{Blais2004}. The connection to the measurement setup is established by a two port microwave transmission line.}
	\label{fig:circuit}
\end{figure}

\begin{figure}
	\centering
	\includegraphics[width=0.9\columnwidth]{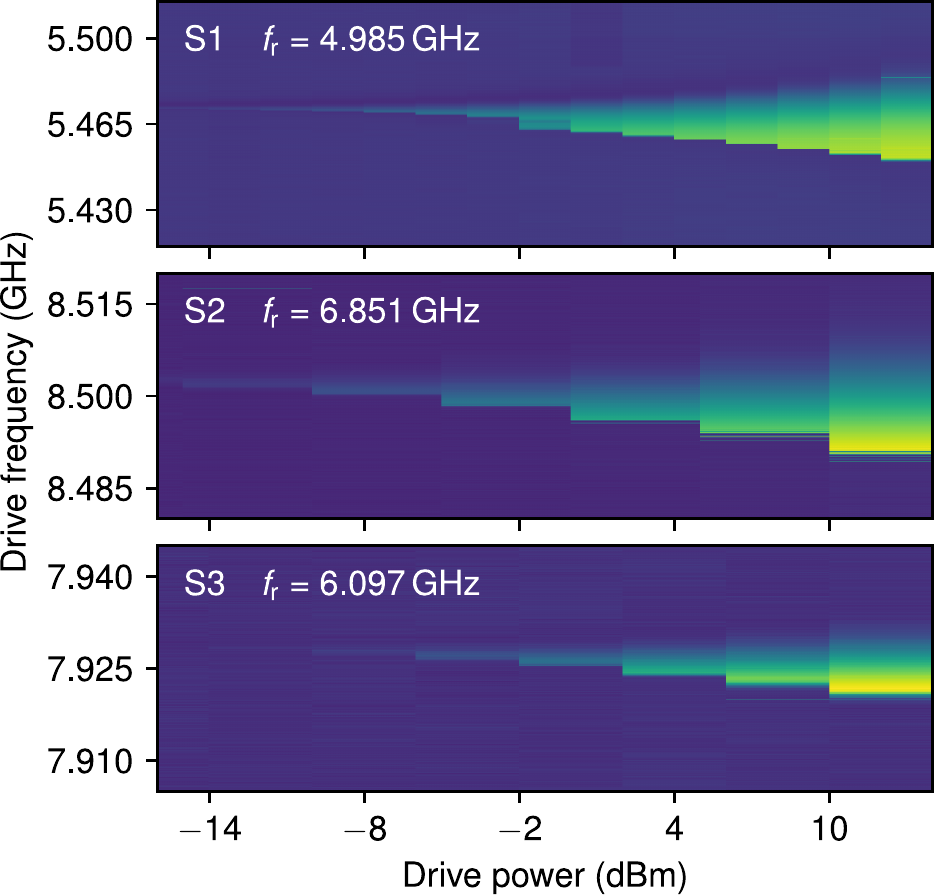}
	\caption{Two-tone spectroscopy of the three samples. If a second microwave drive excites transitions in the sample circuit, the measured amplitude at a fixed frequency in the readout resonance shifts (encoded in the colormap). Due to the line broadening at higher drive power, higher transitions do not show as separate lines.}
	\label{fig:spec}
\end{figure}

\subsection{\label{SN3}{Supplementary Note 3: Numerical modeling}}

In order to understand the interplay of the circuit anharmonicity and potential multi-photon effects due to the microwave driving of the circuit, we simulate the system using the following model. The undriven Hamiltonian of the system is given by
\begin{equation}
\mathcal{H}_0=h(f_{01}+f_\mathrm{s}\sqrt{\epsilon^2/(\epsilon^2+\Delta^2)})\hat{n}-h  f_\mathrm{an}(\hat{n}^2-\hat{n})
\end{equation}
with the bosonic number operator $\hat{n}$. This describes an oscillator with frequency $f_{01}$ and anharmonicity $f_\mathrm{an}$. A value of \SI{1.3}{\MHz} for $f_\mathrm{an}$ is found to agree with other measurements (Fig. 2b in main text). An additional asymmetry extending toward the off resonant regions in the measurement is accounted for by a slight shift $f_\mathrm{s}$ of the main transition in a region defined by a parameter for the width $\epsilon$ and reduced with the detuning $\Delta$ of the drive. This shift might be due to an AC stark effect \cite{Schneider2018}. In the shown result (Fig. 3b in main text) $f_\mathrm{s}$ was set to \SI{-2}{\MHz} and $\epsilon$ to \SI{1}{\MHz}, both adjusted to the off resonant region.

As we are specifically interested in replicating the observed Rabi oscillations, we also need to include the effects of decoherence. We do this by solving the Lindblad-GKS equation \cite{Lindblad1976, Gorini1976},
\begin{equation}
\dot{\rho}=-\ii/\hbar[\mathcal{H},\rho]+\sum_{j=1}^{N}\Gamma_j[L_j \rho L_j^\dagger-1/2\{L_j^\dagger L_j,\rho\}]
\end{equation}
numerically. In this case the total Hamiltonian is comprised of the undriven component and the drive term
\begin{equation}
\mathcal{H} = \mathcal{H}_0 + \mathcal{H}_{\mathrm{drive}}
\end{equation}
where for this circuit we assume that
\begin{equation}
\mathcal{H}_{\mathrm{drive}} = a + a^\dagger
\end{equation}
with the ladder operators defined in the basis of the undriven Hamiltonian such that $\hat{n}=a^\dagger a$. The drive amplitude was adapted to the off-resonant edges with a value of $0.0023$.

For weak microwave driving we could move to a rotating frame to work with a time-independent Hamiltonian. However due to the small anharmonicity we will need to include the possibility of multi-photon transitions. This can be achieved using a Floquet theory \cite{Shirley1965} approach.

The decoherence channels of the Lindblad equation are encoded via the operators $L_j$ and their corresponding rates $\Gamma_j$. The effects of energy dissipation are included via the operator
\begin{equation}
L_1 = a,
\end{equation}
allowing for a decay of one step down. Dephasing is included via
\begin{equation}
L_2 = a^\dagger a.
\end{equation}
To reduce the amount of dynamic parameters, the rates have been fixed to \SI{4}{\micro\second} energy lifetime and \SI{1}{\micro\second} dephasing time, in the order of results from time domain measurements.

To solve the Lindblad equation using Floquet theory, we express the system in an expanded space using the approach detailed in  Ref. \cite{Bain2001}. We then solve the resulting matrix exponential as a function of time, resumming the contributions due to the various Floquet components to obtain the population of the relevant states. From this we calculate the sum of the occupation probability of each state multiplied by a factor of $\sqrt{N}$ to obtain the measurement signal. The additional factor attributes for a reduced dispersive shift of the higher excitations. Convergence of the numerical solution is achieved by increasing both the number of photon manifolds in the Floquet expansion and the number of anharmonic levels in the circuit until the results do not change noticeably. For the calculation of Fig. 3b in the main text, this required 6 Floquet states and 5 circuit levels.

To account for the remaining differences between this model and measurement, several factors can be pointed out. Most importantly, the model Hamiltonian is of a phenomenological form, representing an oscillator with simple anharmonicity. The precise form, however, depends on the exact current phase relation in the system which at this point is not known. Especially at the region in which higher transitions are excited, two further simplifications play a role. It is assumed that both the decay rates and the coupling to the drive will differ between the levels. As there is no way of directly measuring the corresponding rates, we favored a simpler model with fewer free parameters.



\end{document}